\documentclass[final,authoryear,usenatbib,11pt]{elsarticle}

\usepackage{graphicx}
\usepackage{amsmath}

\usepackage{amssymb}

\usepackage{lineno}

\biboptions{square}

\journal{Astroparticle Physics}

\begin{document}

\begin{frontmatter}



\title{Milagro Limits and HAWC Sensitivity for the Rate-Density of Evaporating Primordial Black Holes}


%
\author[MSU,label16]{A. A. Abdo}
\author[MSU]{A. U. Abeysekara}
\author[IF-UNAM]{R. Alfaro}
\author[label2,label17]{B. T. Allen}
\author[UNACH]{C. Alvarez}
\author[UMSNH]{J. D. {\'A}lvarez}
\author[UNACH]{R. Arceo}
\author[UMSNH]{J. C. Arteaga-Vel{\'a}zquez}
\author[UC Santa Cruz,label18]{T. Aune}
\author[MTU]{H. A. Ayala Solares}
% ...

\cortext[corresponding]{Corresponding author: T. N. Ukwatta (tilan.ukwatta@gmail.com)}

\address[MSU]{Department of Physics and Astronomy, Michigan State University, East Lansing, MI, USA}
\address[IF-UNAM]{Instituto de F{\'\i}sica, Universidad Nacional Aut{\'o}noma de M{\'e}xico, Mexico D.F., Mexico}
\address[UNACH]{CEFyMAP, Universidad Aut{\'o}noma de Chiapas, Tuxtla Guti{\'e}rrez, Chiapas, M{\'e}xico}
\address[UMSNH]{Universidad Michoacana de San Nicol{\'a}s de Hidalgo, Morelia, Mexico}
\address[MTU]{Department of Physics, Michigan Technological University, Houghton, MI, USA}
% ...
\address[IGeof-UNAM]{Instituto de Geof{\'\i}sica, Universidad Nacional Aut{\'o}noma de M{\'e}xico, Mexico D.F., Mexico}
% ...
\address[IA-UNAM]{Instituto de Astronom{\'\i}a, Universidad Nacional Aut{\'o}noma de M{\'e}xico, Mexico D.F., Mexico}
% ...
\address[ICN-UNAM]{Instituto de Ciencias Nucleares, Universidad Nacional Aut{\'o}noma de M{\'e}xico, Mexico D.F., Mexico}
% ...
\address[UC Santa Cruz]{Santa Cruz Institute for Particle Physics, University of California, Santa Cruz, Santa Cruz, CA, USA}
% ...
\address[UC Irvine]{Department of Physics and Astronomy, University of California, Irvine, Irvine, CA, USA}
% ...
\address[label2]{Department of Physics and Astronomy, University of California, Irvine, CA 92697}
% ...
\address[label16]{Current address: Operational Evaluation Division, Institute for Defense Analyses, 4850 Mark Center Drive, Alexandria, VA 22311-1882}
\address[label17]{Current address: Harvard-Smithsonian Center for Astrophysics, Cambridge, MA 02138}
\address[label18]{Current address: Department of Physics and Astronomy, University of California, Los Angeles, CA 90095}
% ...

\begin{abstract}
Primordial Black Holes (PBHs) are gravitationally collapsed
objects that may have been created by density fluctuations in the
early universe and could have arbitrarily small masses down to the
Planck scale. Hawking showed that due to quantum
effects, a black hole has a temperature inversely proportional
to its mass and will emit all species of fundamental particles thermally.
PBHs with initial masses of $\sim \, 5.0
\times 10^{14}$ g should be expiring in the present epoch with bursts of
high-energy particles, including gamma radiation in the GeV -- TeV
energy range. The Milagro high energy observatory, which operated
from 2000 to 2008, is sensitive to the high end of the PBH evaporation
gamma-ray spectrum. Due to its large field-of-view, more than 90\% duty
cycle and sensitivity up to 100 TeV gamma rays, the Milagro observatory
is well suited to perform a search for PBH bursts. Based on a
search on the Milagro data, we report new PBH burst rate density upper limits
over a range of PBH observation times. In addition, we report the sensitivity of the Milagro
successor, the High Altitude Water Cherenkov (HAWC) observatory, to PBH evaporation
events.
\end{abstract}


\end{frontmatter}


\section{Introduction}
\label{Intro}

Primordial Black Holes (PBHs) are created from density inhomogeneities in many scenarios of the
early universe~\cite{Carr2010}. The initial mass of a PBH is expected to be roughly equal to or smaller than the horizon or Hubble mass at formation, giving possible PBH masses ranging from that of supermassive black holes down to the Planck mass. PBH production can thus have observable consequences today spanning from the largest scales, for example influencing the development of large-scale structure in the Universe, to the smallest scales, for example enhancing local dark matter clustering. Additionally, PBHs in certain mass ranges may constitute a fraction of the dark matter in the universe~\cite{Carr2010}. For particle physics, the greatest interest is in the radiation directly emitted by a black hole. By evolving an ingoing solution past a gravitationally collapsing object, Hawking showed that a black hole will thermally emit (`evaporate') all available species of fundamental particles~\cite{Hawking1974} with a temperature inversely proportional to the black hole mass. PBHs with an initial mass of $\sim \, 5.0 \times 10^{14}$ g should be expiring now with bursts of high-energy particles, including gamma radiation in the GeV -- TeV energy range~\cite{MacGibbon2008}.

Detection of radiation from a PBH burst would provide valuable insights into the early universe and many areas of physics, as well as confirm the amalgamation of classical thermodynamics with general relativity. Observed PBH radiation will give access to particle physics at energies higher than those which will likely ever be achievable in terrestrial accelerators. Non-detection of PBHs in dedicated searches also gives important information. One of the most important cosmological motivations for PBH searches is to place limits on the initial density fluctuation spectrum of the early universe. In particular, PBHs can form from the quantum fluctuations associated with many types of inflationary scenarios~\cite{Carr2005}. Other PBH formation mechanisms include those associated with cosmological phase transitions, topological defects or an epoch of low pressure (soft equation of state) in the early universe (for a review see ~\cite{Carr2010}).

Evaporating PBHs are candidate sources for gamma-ray telescopes, as they produce short bursts of gamma rays.
While most gamma-ray bursts (GRBs) are generally thought to be produced by the collapse of
massive stars (long duration GRBs) or the merger of compact
objects (short duration GRBs) at cosmological distances~\cite{Gehrels2012},
some short duration GRBs
show behavior, such as large offsets from the host galaxy or
anisotropic sky distribution, that may indicate a more local origin.
Models of PBH evaporation based on Standard Model physics predict a unique TeV gamma-ray spectrum \cite{Halzen1991}.

Various detectors have searched for PBH events using direct and
indirect methods. These methods probe the PBH distribution on various
distance scales. One can probe the PBH density on the
cosmological scale using the 100 MeV extragalactic gamma-ray
background, which produces a limit on the corresponding cosmological average PBH burst rate density of $< \, 10^{-6}$ ${\rm pc^{-3} yr^{-1}}$~\cite{Carr2010, PageHawking1976}. On the galactic scale, if PBHs are clustered in the Galaxy, we would expect to see an enhancement in the local PBH density and anisotropy in the 100 MeV gamma-ray measurements. Indeed, such an anisotropy has been detected and results in a corresponding Galactic PBH burst limit of $<$ 0.42 ${\rm pc^{-3} yr^{-1}}$~\cite{Wright1996}. On the kiloparsec scale, the Galactic antiproton background can be used to give a PBH burst limit of $<$ 0.0012 ${\rm pc^{-3} yr^{-1}}$~\cite{Abe2012}. However, the antiproton-derived limit depends on the assumed PBH distribution within the Galaxy and the propagation of antiprotons through the Galaxy, as well as the production and propagation of the secondary antiproton component produced by interactions of cosmic-ray
nuclei with the interstellar gas. On the parsec scale, the PBH burst limits are directly set by searches for the detection of individual bursting PBHs and are independent of assumptions of PBH clustering. The best direct search limits come from the Very High Energy (VHE) searches conducted with Imaging Air Cherenkov Telescopes (IACTs) and Extensive Air Shower (EAS) arrays. On the parsec scale, the current best PBH limit from direct searches is $<$ $1.4 \times 10^{4}$ ${\rm pc^{-3} yr^{-1}}$~\cite{Glicenstein2013}. 

%
\section{Milagro and HAWC Observatories}
\label{milagro_hawc}

In this paper, we present new PBH burst limits based on the direct search method using
the data from the Milagro observatory. These limits are obtained
assuming the standard model of Hawking radiation and particle
physics~\cite{MacGibbon1990, Halzen1991}. Milagro was a water Cherenkov
gamma-ray observatory (EAS type) sensitive to gamma rays in the energy
range $\sim$ 50 GeV to 100 TeV. The observatory was located near Los Alamos,
New Mexico, USA at latitude $35.9^{\circ}$ north, longitude $106.7^{\circ}$ west
and an altitude of 2630 m, and was operational from 2000 to 2008~\cite{Atkins2000}. The
Milagro detector had two components: a central rectangular 60 m $\times$ 80 m $\times$ 7 m
reservoir filled with purified water and an array of 175 smaller outrigger (OR) tanks
distributed over an area of 200 m $\times$ 200 m surrounding the reservoir. The reservoir was light-tight and instrumented with two layers of 8-inch photomultiplier tubes (PMTs).
The top layer consisted of 450 PMTs (the `air-shower' layer) 1.5 m below the water surface
and the bottom layer had 273 PMTs (the `muon' layer) 6 m below the surface.
Each outrigger tank contained one PMT. The observatory detected VHE gamma rays
by detecting the Cherenkov light generated by the secondary particles from the gamma-ray-induced air shower as the secondary particles passed through the water. Various components of the detector were used to measure the direction of the gamma ray and
to reduce the background due to hadron-induced showers.
Because of its large field-of-view, $\sim$2 sr, and high duty
cycle, over 90\%, Milagro was well suited to search for burst
emission from PBHs.

In this paper we also present the sensitivity of the High Altitude Water Cherenkov (HAWC)
observatory to PBH bursts. HAWC, the successor to Milagro, is the next generation
VHE observatory presently under construction at Sierra Negra,
Mexico at an altitude of 4100m. HAWC will consist of 300 water tanks, each 7.3 m in diameter
and 4.5 m deep. Each tank will house three 8-inch PMTs (reused from Milagro) and
one 10-inch PMT~\cite{HAWC_TeV_Sensi_2013}. The PMTs will detect Cherenkov light from secondary particles created in extensive air showers induced by VHE gamma rays of energy in
the range $\sim$50 GeV to 100 TeV.
HAWC has two data acquisition (DAQ) systems: the main DAQ and the scaler DAQ.
The main DAQ system measures the arrival direction and energy of the high-energy
gamma-rays by timing the arrival of particles on the ground. The direction of the
original primary particle may be resolvable with an error of between 0.1 and 2.0 degrees
depending on its energy and location in the sky. The scaler DAQ system counts the number
of hits in each PMT, allowing a search for excesses over background noise.
The scaler DAQ system is more sensitive to lower energy air showers than the main DAQ system.
HAWC has a large field-of-view (1.8 sr corresponding to 1/7 th of the sky) and will have a high duty cycle of greater than 90\%. Thus, HAWC will be able to observe high-energy emission
from gamma-ray transients~\cite{HAWC_GRB_Sensi_2012}.

\section{Methodology}\label{Methodology}

\subsection{Primordial Black Hole Burst Spectrum}

The properties of the final burst of radiation from a PBH depend on the physics
governing the production and decay of high-energy particles.
As the black hole evaporates, it loses mass and hence its
temperature and the number of particle species that it emits
increase until the end of its evaporation lifetime. In the Standard Evaporation Model (SEM)~\cite{MacGibbon1990, MacGibbon1991}, a PBH
should directly radiate those fundamental particles whose Compton wavelengths are on the order of the size of the black hole. When the black hole temperature exceeds
the Quantum Chromodynamics (QCD) confinement scale (250--300 MeV),
quarks and gluons should be directly emitted by the black hole~\cite{PageHawking1976, MacGibbon1990}. The quarks and gluons will then fragment and hadronize as they stream away from the black hole, analogous to the jets seen in terrestrial accelerators~\cite{MacGibbon1990, MacGibbon2008}. On astrophysical timescales, the jet products will decay into photons, neutrinos, electrons, positrons, protons and anti-protons.

Detailed studies using the SEM to simulate the particle spectra from
black holes with temperatures of $1 - 100$ GeV have shown that the
gamma-ray spectrum is dominated by the photons produced by neutral pion decay in the
Hawking-radiated QCD jets and is broadly peaked at photon energies of
$\sim$100 MeV. The photons which are directly radiated (not the result
of decay of other primary particles) are visible as a much smaller peak at a much higher
photon energy proportional to the black hole temperature~\cite{MacGibbon1990}.
As the evaporation proceeds to higher temperatures, the greater will be the number of emitted
fundamental particle degrees of freedom. As new particle degrees of freedom
become available, the luminosity of the burst will increase.
The energy spectrum is determined by the correct high energy particle physics model. In this work,
we will assume the SEM as our emission and particle physics model.

\begin{figure}
\centering
\includegraphics[width=0.8\textwidth]{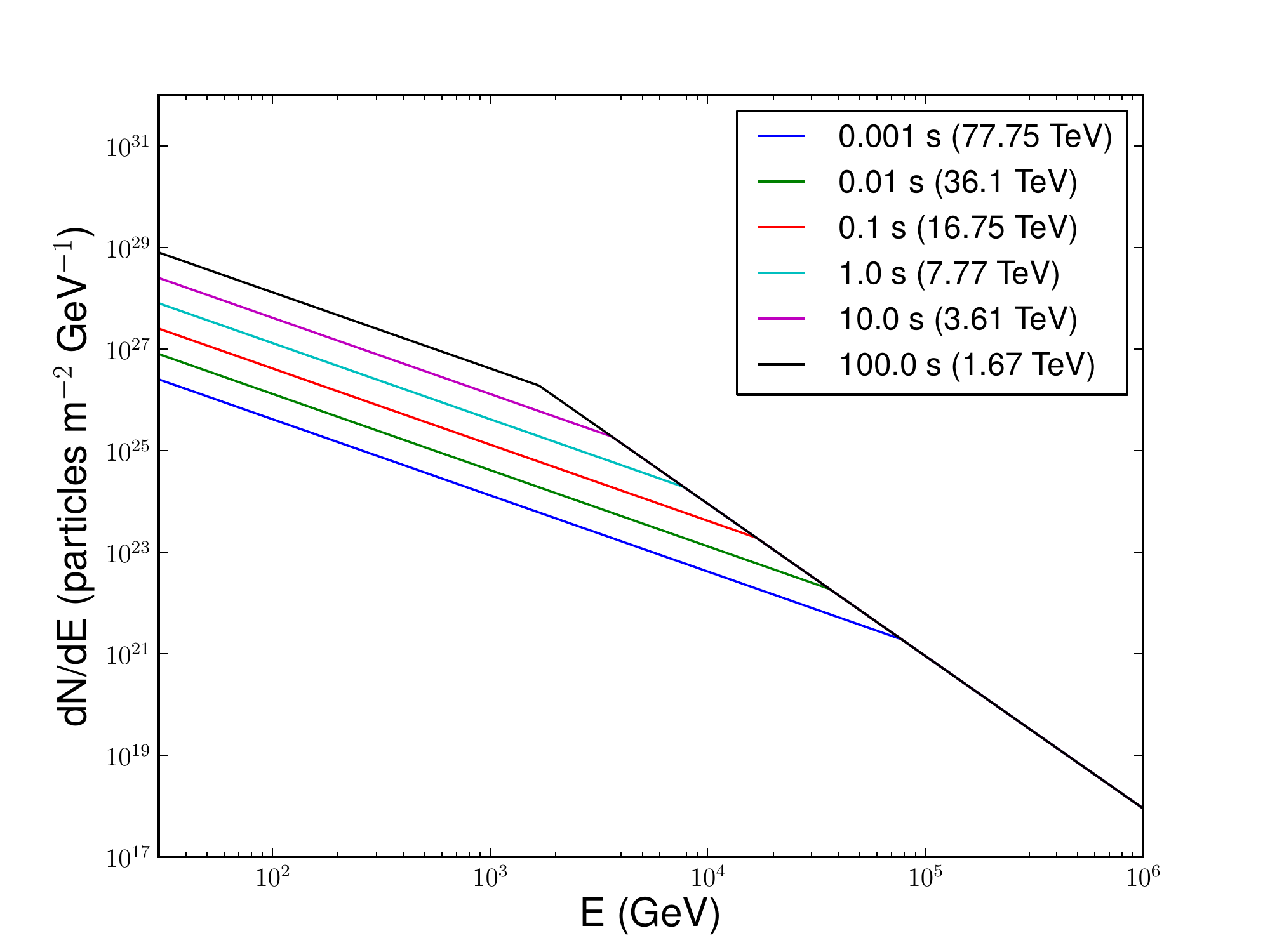}
\caption{Time-integrated gamma-ray spectrum over various PBH remaining lifetimes using the parametrization of Equation \ref{photonEq}. The black hole
temperature at the start of observation is shown in parentheses.}\label{pbh_spectrum}
\end{figure}

For temperatures well below the Planck temperature, the temperature ($T$) of a black hole
can be expressed in terms of the remaining evaporation lifetime ($\tau$) of the black hole
(that is, the time left until the black hole stops evaporating) as follows~\cite{MacGibbon1991,Petkov2008}:
\begin{equation} \label{tempEq}
T \simeq \bigg[4.8 \times 10^{11} \, \bigg(\frac{\rm{1 sec}}{\tau}\bigg) \bigg]^{1/3}\,\,\,\rm GeV.
\end{equation}
The emission rate increases as the black hole shrinks and therefore the remaining
evaporation lifetime decreases as $T$ increases. For black holes with
temperatures greater than several GeV at the start of the observation,
the time integrated photon flux can be parameterized as~\cite{Petkov2008}
\begin{equation} \label{photonEq}
\frac{dN}{dE} \approx 9 \times 10^{35} {\,\, \rm particles \,\, m^{-2} GeV^{-1}}
\begin{cases}
\big(\frac{1 GeV}{T}\big)^{3/2}\big(\frac{1 GeV}{E}\big)^{3/2},\,\,\,\,E<T \\
\big(\frac{1 GeV}{E}\big)^{3},\,\,\,\,E\ge T
\end{cases}
\end{equation}
for gamma ray energies $E \gtrsim$ 10 GeV. The $E^{-3}$ fall off at $E \ge T$ comes
from the $\tau \propto T^{-3}$ dependence in Equation~\ref{tempEq}, which is less steep
than the high energy exponential tail in the instantaneous Hawking spectrum at
each $T$~\cite{Petkov2008}. Figure~\ref{pbh_spectrum} shows the resulting gamma-ray spectrum
for various PBH remaining lifetimes ranging from 0.001 s to 100 s.”

\subsection{Detectable Volume Estimation}
In order to calculate the upper limits on the PBH burst rate density, it is
necessary to calculate the PBH detectable volume for a given detector.
In general, the expected number of photons detectable by an
observatory on the Earth's surface from a PBH burst of duration $\tau$
at a non-cosmological distance $r$ and zenith angle $\theta$ is
\begin{equation} \label{countsEq}
\mu(r, \theta, \tau) = \frac{(1-f)}{4 \pi r^2} \int_{E_1}^{E_2} \,\frac{dN(\tau)}{dE}\, A(E,
\theta)\,dE
\end{equation}
where $f$ is the dead time fraction of the detector and $dN/dE$ is
the black hole gamma ray spectrum integrated from remaining time $\tau$ to 0.
The values $E_1$ and $E_2$ correspond to the lower and upper bounds respectively
of the energy range searched and $A(E,\theta)$ is the
effective area of the detector as a function of photon energy and
zenith angle. Typically the function $A(E, \theta)$ is obtained
from a simulation of the detector. For Milagro and HAWC, the dependence
of $A(E, \theta)$ on the zenith angle is usually given
in discrete bands (represented by $\theta_i$). We will define these bands
specifically for a given observatory in Section~\ref{Results}.

The background cosmic-ray flux at the Earth's surface is much higher
than the background gamma-ray flux. Most events detected by EAS arrays
such as Milagro or HAWC are air showers induced by cosmic rays.
To search for the emission from a PBH burst one needs to look for
an excess that cannot be explained by statistical fluctuations of the background.

In this paper, we estimate $\mu_{\circ}(\theta_i, \tau)$, the minimum number of counts
needed for a detection for different burst durations $\tau$, by finding the number
of counts required over the background for a $5 \sigma$ significance (after
correcting for multiple trials). First, we calculate the background rates ($R(\theta_i)$)
utilizing a Monte Carlo simulation (for HAWC) or actual data (for Milagro). The background rate depends on the spatial bin-size, burst duration and background rejection
parameters, as well as the zenith angle band $\theta_i$. In section~\ref{Results}, we optimize these parameters to minimize the
background rates and maximize the sensitivity. Using the background rates, we then find the
$\mu_{\circ}(\theta_i, \tau)$ values required for a 50\% probability of
detecting a $5\sigma$ excess after a given number of trials, based on the Poisson
fluctuations of the signal around $\mu_{\circ}(\theta_i, \tau)$.

We define a 5$\sigma$ detection after correction for $N_t$ trials
to be the number of counts $n$ which would have a Poisson probability
$P$ corresponding to a Bonferroni corrected p-value $p_c$  given by
\begin{equation} \label{stat1}
p_c = p_o /N_t = P(\geq n|n_{\rm bk}).
\end{equation}
Here, $p_0 = 2.3 \times 10^{-7}$ is the p-value corresponding to 5$\sigma$,
$n_{\rm bk}=\tau R(\theta_i)$ is the number of background counts
expected over the burst duration $\tau$, and $P(\geq n|n_{\rm bk})$ denotes the
Poisson probability of getting $n$ or more counts when the Poisson mean is $n_{\rm bk}$.
We take the value of $\mu_{\circ}(\theta_i, \tau)$ to be the amount of expected signal which
would satisfy this criterion 50\% of the time. We find $\mu_{\circ}(\theta_i, \tau)$ by estimating the Poisson probability $P$ of finding at least $n$ counts to be 50\% according to the relation
\begin{equation} \label{stat2}
P(\geq n | (n_{\rm bk} + \mu_{\circ}(\theta_i, \tau))) = 0.5.
\end{equation}

By equating the values of $\mu_{\circ}(\theta_i, \tau)$ found from Equation~\ref{stat2} to $\mu(r, \theta_i, \tau)$ in Equation~\ref{countsEq} and solving for $r$, we calculate the maximum distance from which a PBH burst could be detected by the high-energy observatory,
\begin{equation} \label{distanceEq}
r_{\rm max}(\theta_i, \tau) = \sqrt{ \frac{(1-f)}{4 \pi \mu_{\circ}(\theta_i, \tau)} \int_{E_1}^{E_2} \,\frac{dN(\tau)}{dE}\, A(E,
\theta_i)\,dE}
\end{equation}
for various zenith angle bands $\theta_i$ and burst durations $\tau$. Denoting the effective
field-of-view of the detector for a given zenith band by
\begin{equation} \label{fov}
{\rm FOV_{\rm eff}}(\theta_{i}) = 2 \pi (\cos \theta_{i,\,\rm min}-\cos \theta_{i,\,\rm max})\ {\rm sr}
\end{equation}
where $\theta_{i,\, \rm min}$ and $\theta_{i,\, \rm max}$ are the minimum and maximum zenith angles in band $i$,
the detectable volume is then
\begin{equation} \label{volueEq1}
V(\tau) = \sum_{i} V(\theta_{ i}, \tau) = \frac{4}{3} \pi \sum_{i} r_{\rm max}^3(\theta_{ i}, \tau) \frac{{\rm FOV_{\rm eff}}(\theta_{i})}{4\pi}.
\end{equation}

\subsection{Upper Limit Estimation}

If the PBHs are uniformly distributed in the solar neighborhood, the
X\% confidence level upper limit ($UL_{X}$) on the rate density of
PBHs bursts (that is, the number of bursts occurring locally per unit volume per unit time) can be estimated as
\begin{equation}\label{ulX}
UL_{X} = \frac{m}{V S}
\end{equation}
if at the X\% confidence level the detector observes zero bursts. Here $V$ is the effective detectable volume from which a PBH can be detected, $S$ is the total search duration and $m$ is the upper limit on the expected number of PBH bursts given that at the X\% confidence level zero bursts are observed at Earth. Note that for Poisson fluctuations $P(0|m)=1-X$ and so
$m = \ln (1/(1-X))$. If $X=99\%$ and $m=\ln 100 \approx 4.6$, the upper limit on the PBH burst rate density will be

\begin{equation}\label{ul99}
UL_{99} = \frac{4.6}{V S}.
\end{equation}

\section{Results}\label{Results}

\subsection{Milagro Limits on the Rate-Density of PBH Bursts}

During the early days of its operation, Milagro had lower angular
resolution prior to the addition of the outrigger array and used a
triggering system that did not accept many of the low energy events.
Thus, for this search we used the last five years of Milagro data: specifically from 03/01/2003 to 03/01/2008. (Due to various detector-related issues, 7\% of the data taken
during this period was also not used.) Selection cuts were made to increase the quality of the data searched. Reconstructed events which have a predicted angular reconstruction error greater than 2$^\circ$ were rejected. (This corresponds to $nfit >$20 where $nfit$ is the number of PMTs
participating in the reconstruction of the shower.) The maximum zenith angle
used was 45$^\circ$ and the best limits were obtained with no gamma-hadron separation cut applied, because such a cut also strongly lowered the Milagro photon effective area at energies below 1 TeV. Overall, our analysis utilized 1673 days (4.58 years)
worth of good data, amounting to $\sim 93$\% of the total Milagro data
collected during the five year period (neglecting the dead time).
The Milagro search and its optimization presented here are described in
further detail in~\cite{Vlasios_Thesis_2008}.

We performed a blind search (that is, a search utilizing no external triggers) for burst durations ranging from 250 $\mu$s to 6 minutes. First we created skymaps for overlapping time intervals, each offset by 10\% of the pre-set burst duration. We then spatially binned the skymap and searched for locations with significant excess over background in the Milagro data. The optimum bin-size was determined using a Monte-Carlo simulation and varied with the pre-set burst duration. For short durations the optimum bin size was of order $\sim$1.8$^\circ$ and for long durations it was $\sim$0.8$^\circ$: for short durations, Milagro was signal-limited requiring a larger bin-size to accumulate more signal; for long durations, Milagro became background-dominated,
requiring a more restricted bin-size to reduce background contamination.

No statistically significant (5$\sigma$) event was observed over the 4.58 years of data. Proceeding on the basis of null detection, we calculated the upper limits on the PBH burst rate density following the methodology described in Section~\ref{Methodology}. For Milagro, we parameterized the effective area as $A(E, \theta_i) = 10^{a_i(\log E)^2 + b_i \log E + c_i} \,\,\, \rm m^2$ with the parameters $a_i$, $b_i$ and $c_i$ given in Table~\ref{para_table_milagro} for three zenith angle bands.
Figure~\ref{effective_area} shows the Milagro effective area curves for the
selected three zenith angle bands.
\begin{table}[h]
\begin{center}
\begin{tabular}{|c|c|c|c|}
\hline Zenith Angle Band $\theta_i$ & $a_i$ & $b_i$ & $c_i$   \\ \hline
0$^{\circ}$ - 15$^{\circ}$  ($\theta_1$) & -0.4933 & 4.7736 & -2.4272 \\ \hline
15$^{\circ}$ - 30$^{\circ}$ ($\theta_2$) & -0.5037 & 5.0102 & -3.4015 \\ \hline
30$^{\circ}$ - 45$^{\circ}$ ($\theta_3$) & -0.4273 & 4.7931 & -4.3030 \\ \hline
\end{tabular}
\caption{Milagro effective area parametrization for various zenith angle bands.}
\label{para_table_milagro}
\end{center}
\end{table}

We utilized a Monte Carlo simulation to calculate
the $\mu_{\circ}(\theta_i, \tau)$ values for various burst durations.
Because the trials in our search were not independent,
we took $N_t$ in Equation~\ref{stat1} to be the effective number of independent trials calculated using the method
described in~\cite{Vlasios_Thesis_2008}. The effective number of independent trials
ranged from $\sim$0.1\% to $\sim$40\% of the total number of trials depending on the search duration, with the shorter search durations having the lower fraction of effective trials.
The dependence of the limit on the estimated number of independent trials is quite mild
($\sim N_t^{0.018}$) so that varying the estimated trials by 3 orders of magnitude produces less than 15\% change in the limit. The resulting $\mu_{\circ}(\theta_i, \tau)$ values are listed in Table~\ref{mu_limit_table}. These $\mu_{\circ}(\theta_i, \tau)$ values and the Milagro effective area parameterizations were then used to derive the maximum
detectable distance of a PBH burst, $r_{\rm max}(\theta_i, \tau)$, using Equation~\ref{distanceEq} assuming an energy range of
$E_1$=50 GeV and $E_2$=100 TeV and a deadtime of 7\%.
The derived $r_{\rm max}(\theta_i, \tau)$ values were then used to calculate the
effective volume that was probed by the Milagro observatory. Using Equation~\ref{ul99}, we calculated 99\% upper limits for various PBH remaining lifetimes and the effective total search period of 4.58 years. Our results are shown in
Table~\ref{mu_limit_table} and in Figure~\ref{pbh_limits}. According to our results,
Milagro is most sensitive to burst durations of about 1 s. For shorter
durations, the Milagro data is starved for signal photons; for longer durations, the background tends to dominate the signal. We note that Milagro has a systematic flux uncertainty of $\sim$30\% ~\cite{MilagroCrab2012} which translates into an $\sim$50\% uncertainty in our calculated limit (shown as a pink shaded
band in Figure~\ref{pbh_limits}).

\begin{table}[h]
\begin{center}
\begin{tabular}{|l|c|c|}
\hline Burst Duration $\tau$ (s) & $\mu_{\circ, \tau}$ & $UL_{99}$ ($\rm pc^{-3} yr^{-1}$) \\ \hline
0.001 & 11 & 3.1 $\times 10^5$ \\ \hline
0.01 & 16 & 1.2 $\times 10^5$ \\ \hline
0.1 & 22 & 5.4 $\times 10^4$ \\ \hline
1.0 & 35 &  3.6 $\times 10^4$ \\ \hline
10.0 & 65 &  3.8 $\times 10^4$ \\ \hline
100.0 & 150 &  6.9 $\times 10^4$ \\ \hline
\end{tabular}
\caption{Counts ($\mu_{\circ}(\tau)$) needed over the background for a $5\sigma$ detection with 99\% probability and calculated 99\% confidence upper limits ($UL_{99}$) for various burst durations ($\tau$) for Milagro. }
\label{mu_limit_table}
\end{center}
\end{table}


\subsection{Improved HAWC Sensitivity to the Rate Density of PBH Bursts}

\begin{figure}
\centering
\includegraphics[width=0.9\textwidth]{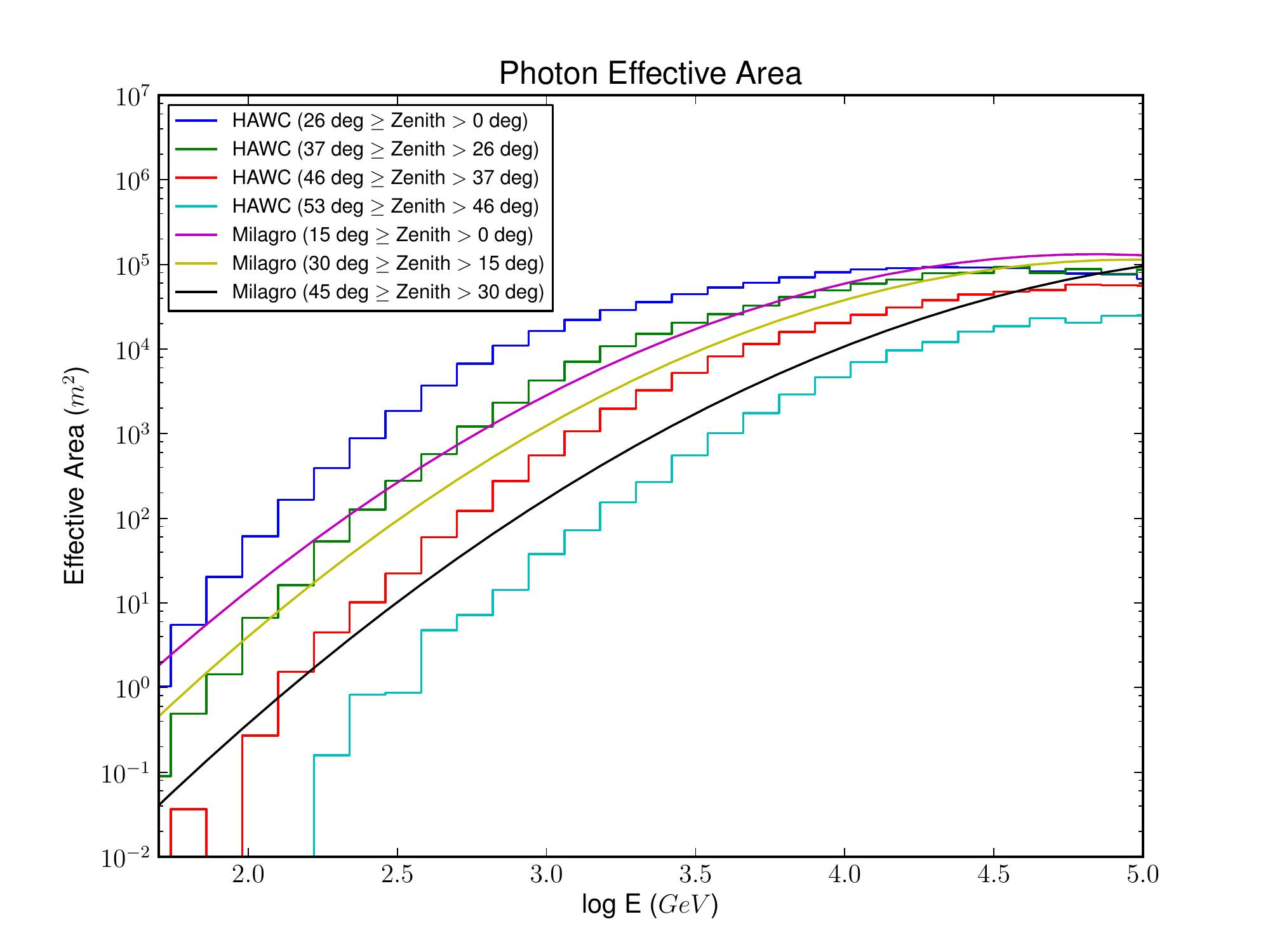}
\caption{Effective Area for photon detection for HAWC and Milagro as a function of energy.
The Milagro effective area curves use $nHit>$50, $DelAngle <$1.5 deg and
no gamma-hadron separation cuts. The HAWC effective area curves use $nHit >$100, $DelAngle <$0.8 deg and $nHit/CxPE40>$ 7. The HAWC cuts are optimized for the PBH spectrum and utilize an $nHit$ cut that is well
above the intrinsic threshold. This and the fact that Milagro used no gamma-hadron cut result in an effective area for Milagro which is larger than for HAWC at low energies. However,
HAWC has superior sensitivity in the PBH search.}\label{effective_area}
\end{figure}

Milagro's successor HAWC is located at a higher altitude and features a better
detector design, allowing for superior sensitivity to PBH bursts. In this
section, we apply our methodology to estimate the HAWC
sensitivity to PBH bursts. In our calculations, all relevant characteristics of the HAWC detector are encoded in the effective area. We calculate the effective area using a Monte Carlo simulation which models the interaction of photons and
cosmic rays in the atmosphere and the response of the detector to the
extensive air showers generated by these particles. The effective area is
then defined as the ratio of the number of events that satisfies a given set of
cuts to the total number of events multiplied by the total throw area
of the Monte Carlo simulation. In our case the Monte Carlo throw area is a
circle of 1000m radius. The cuts are comprised of a trigger
cut, an angle cut and a gamma-hadron separation cut. For the trigger,
HAWC will use events with $nHit$, the number of PMTs hit by the air shower,
greater than a certain value. The angle cut is employed to specify the direction
of the photons and is a measure of HAWC's angular resolution. In the simulated
events we use an angular parameter $DelAngle$ which is the difference between the
true location of the particle in the sky and the reconstructed sky location of the particle.
This is a proxy for the angular search bin-size.
Because the background is predominantly protons, the angle cut was not used to calculate the
HAWC effective area for protons and the gamma-hadron separation cut was used to reduce
background events. The standard gamma-hadron separation parameter for HAWC is called
compactness and is defined as $nHit/CxPE40$ where $CxPE40$ is the number of photoelectrons
recorded in the strongest hit PMT outside a 40m radius from the reconstructed
shower core. The shower core represents the location on the ground where the original particle
would have hit had it not interacted with the atmosphere. Figure~\ref{effective_area}
shows the HAWC effective area $A(E, \theta_i)$ for photons as a function of photon energy and zenith angle band using the optimum cuts for a 10 s burst search ($nHit >$100, $DelAngle <$0.8 deg, and
with $nHit/CxPE40>$ 7).

In order to estimate the background rate $R_b(\theta, \xi)$ where $\xi$ is a measure of the spatial resolution of HAWC, we used the ATIC cosmic ray
spectrum given in Reference~\cite{atic2006}
\begin{equation}\label{atic_spectrum}
\frac{dN_p}{dE} = 7900 \times \bigg(\frac{E}{\rm 1 GeV}\bigg)^{-2.65} \,\,\,\,\, \rm particles \,\, m^{-2} s^{-1} sr^{-1} GeV^{-1}
\end{equation}

and convolved it with the HAWC proton effective area for a given zenith angle band,

\begin{equation}\label{background_rate}
R_b(\theta_i, \xi) = \int_{E_1}^{E_2} \,\frac{dN_p}{dE}\, A_p(E, \theta_i)\,dE \times 2 \pi (1.0 - \cos(\xi)) \times 1.2
\end{equation}
In Equation~\ref{background_rate}, the $2 \pi (1.0 - \cos(\xi))$ term represents the patch in the sky that corresponds to the spatial resolution of HAWC in steradians. In our case $\xi = DelAngle$. The factor 1.2 is a correction factor incorporating the other particle species in the cosmic ray background.

\begin{table}[h]
\begin{center}
\begin{tabular}{|l|l|}
\hline Parameter & Values  \\ \hline
$nHit$ $\ge$ & 30, 70, 100, 150, 200, 250 \\ \hline
$DelAngle$ $\le$ & 0.1$^\circ$, 0.3$^\circ$, 0.8$^\circ$, 1.0$^\circ$, 3.0$^\circ$, 8.0$^\circ$\\ \hline
Compactness $\ge$ & 1, 2, 3, 4, 5, 6, 7, 8, 9, 10 \\ \hline
\end{tabular}
\caption{Various parameter cuts used for the simple parameter search.}
\label{para_search}
\end{center}
\end{table}

Because we seek the sensitivity in the case where there is no prior knowledge of the burst location, we need to take into account the number of trials performed for the search.
For example, if the HAWC field-of-view (1.8 sr) is divided into spatial bins of radius 0.7$^\circ$, then there will be approximately 10$^4$ spatial bins (trials) per time bin searched. The optimum spatial bin-size depends on the search duration, the trigger criteria, and the value of the gamma-hadron separation parameter.
The number of time bins is estimated by dividing the total search period (estimated as 5 years for HAWC) by the burst duration $\tau$. Thus the total number of trials depends on the burst duration $\tau$, the optimal spatial bin-size $DelAngle$, the trigger criterion $nHit$ and the value of the compactness parameter. In order to find the optimum set of cuts, we performed a simple parameter search and identified the set of values which give the best PBH limit according to the method described in Section~\ref{Methodology}.

\begin{table}[h]
\begin{center}
\begin{tabular}{|l|c|c|c|}
\hline Duration $\tau$ (s) & nHit & DelAngle (deg) & Compactness \\ \hline
0.001 & 30 & 3.0 & 3  \\ \hline
0.01 & 30 & 3.0 & 3  \\ \hline
0.1 & 70 & 1.0 & 4  \\ \hline
1.0 & 70 & 1.0 & 5  \\ \hline
10.0 & 100 & 0.8 & 7  \\ \hline
100.0 & 100 & 0.8 & 7  \\ \hline
\end{tabular}
\caption{Optimized cuts for various burst durations.}
\label{para_table_hawc}
\end{center}
\end{table}

For burst durations ranging from 0.001 s to 100 s, we performed cuts on all parameter combinations given in Table~\ref{para_search} on the Monte Carlo output and calculated corresponding effective areas for photons and protons. Using Equations~\ref{atic_spectrum} and \ref{background_rate}, we then calculated the
background rate and the background number density $n_{\rm bk}(\theta_i, \xi)=\tau R_b(\theta_i, \xi)$ (see Equation~\ref{stat1}) which depends on the zenith angle band and the spatial resolution. As remarked earlier, the effective number of independent trials differs for each parameter combination. Taking this into account, we have calculated
the $\mu_{\circ}(\theta_i, \tau)$ values corresponding to burst durations ranging from 0.001
s to 100 s for various zenith angle bands. These $\mu_{\circ}(\theta_i, \tau)$ values and the effective area for photons are then inserted into Equation~\ref{distanceEq} (with $E_1$=50 GeV and $E_2$=100 TeV) and the maximum distances $r_{\rm max}(\theta_i, \tau)$ at which a PBH burst can be detected by the HAWC observatory is calculated for various burst durations $\tau$ assuming negligible dead time. Using these $r_{\rm max}(\theta_i, \tau)$ values and Equation~\ref{volueEq1}, we have determined the effective detectable volume $V(\tau)$ for each burst duration. The PBH limit for each parameter combination was calculated using Equation~\ref{ul99} and the set of cuts that gives the best limit for a given burst duration was selected. The resulting optimized parameter cut values are given in Table~\ref{para_table_hawc}. The corresponding values of $\mu_{\circ}(\theta_i, \tau)$ for a 5$\sigma$ detection are given in Table~\ref{mu_table} with the associated background counts and number of trials factor.

\begin{table}[h]
\begin{center}
\scriptsize
\begin{tabular}{|l|c|c|c|c|}
\hline Duration $\tau$ (s) & Zenith Angle Band $\theta_i$ & Number of Trials & Bgnd. Counts $n_{\rm bk}$ & $\mu_{\circ}(\theta_i, \tau)$ \\ \hline
0.001 & 0$^{\circ}$ - 26$^{\circ}$ (${\theta}_1$) &  $3.3 \times 10^{13}$  & 0.0637 & 10.6  \\ \hline
0.001 & 26$^{\circ}$ - 37$^{\circ}$ (${\theta}_2$) &  $3.3 \times 10^{13}$  & 0.0240 & 8.6  \\ \hline
0.001 & 37$^{\circ}$ - 46$^{\circ}$ (${\theta}_3$) &  $3.3 \times 10^{13}$  & 0.0083 & 7.7  \\ \hline
0.001 & 46$^{\circ}$ - 53$^{\circ}$ (${\theta}_4$) &  $3.3 \times 10^{13}$  & 0.0026 & 6.7  \\ \hline
0.01 & 0$^{\circ}$ - 26$^{\circ}$ (${\theta}_1$) &  $3.3 \times 10^{12}$  & 0.6372 & 17.0  \\ \hline
0.01 & 26$^{\circ}$ - 37$^{\circ}$ (${\theta}_2$) &  $3.3 \times 10^{12}$  & 0.2397 & 13.4  \\ \hline
0.01 & 37$^{\circ}$ - 46$^{\circ}$ (${\theta}_3$) &  $3.3 \times 10^{12}$  & 0.0832 & 10.6  \\ \hline
0.01 & 46$^{\circ}$ - 53$^{\circ}$ (${\theta}_4$) &  $3.3 \times 10^{12}$  & 0.0256 & 8.6  \\ \hline
0.1 & 0$^{\circ}$ - 26$^{\circ}$ (${\theta}_1$) &  $3.0 \times 10^{12}$  & 0.1355 & 11.5  \\ \hline
0.1 & 26$^{\circ}$ - 37$^{\circ}$ (${\theta}_2$) &  $3.0 \times 10^{12}$  & 0.0456 & 9.6  \\ \hline
0.1 & 37$^{\circ}$ - 46$^{\circ}$ (${\theta}_3$) &  $3.0 \times 10^{12}$  & 0.0144 & 7.7  \\ \hline
0.1 & 46$^{\circ}$ - 53$^{\circ}$ (${\theta}_4$) &  $3.0 \times 10^{12}$  & 0.0036 & 6.7  \\ \hline
1.0 & 0$^{\circ}$ - 26$^{\circ}$ (${\theta}_1$) &  $3.0 \times 10^{11}$  & 1.0481 & 18.6  \\ \hline
1.0 & 26$^{\circ}$ - 37$^{\circ}$ (${\theta}_2$) &  $3.0 \times 10^{11}$  & 0.3422 & 14.3  \\ \hline
1.0 & 37$^{\circ}$ - 46$^{\circ}$ (${\theta}_3$) &  $3.0 \times 10^{11}$  & 0.1055 & 10.6  \\ \hline
1.0 & 46$^{\circ}$ - 53$^{\circ}$ (${\theta}_4$) &  $3.0 \times 10^{11}$  & 0.0251 & 8.6  \\ \hline
10.0 & 0$^{\circ}$ - 26$^{\circ}$ (${\theta}_1$) &  $4.6 \times 10^{10}$  & 2.4405 & 23.2  \\ \hline
10.0 & 26$^{\circ}$ - 37$^{\circ}$ (${\theta}_2$) &  $4.6 \times 10^{10}$  & 0.7039 & 16.0  \\ \hline
10.0 & 37$^{\circ}$ - 46$^{\circ}$ (${\theta}_3$) &  $4.6 \times 10^{10}$  & 0.1912 & 11.5  \\ \hline
10.0 & 46$^{\circ}$ - 53$^{\circ}$ (${\theta}_4$) &  $4.6 \times 10^{10}$  & 0.0451 & 8.6  \\ \hline
100.0 & 0$^{\circ}$ - 26$^{\circ}$ (${\theta}_1$) &  $4.6 \times 10^{9}$  & 24.4049 & 51.3  \\ \hline
100.0 & 26$^{\circ}$ - 37$^{\circ}$ (${\theta}_2$) &  $4.6 \times 10^{9}$  & 7.0394 & 31.6  \\ \hline
100.0 & 37$^{\circ}$ - 46$^{\circ}$ (${\theta}_3$) &  $4.6 \times 10^{9}$  & 1.9118 & 20.8  \\ \hline
100.0 & 46$^{\circ}$ - 53$^{\circ}$ (${\theta}_4$) &  $4.6 \times 10^{9}$  & 0.4513 & 14.2  \\ \hline
\end{tabular}
\caption{Counts $\mu_{\circ}(\theta_i, \tau)$ needed over the background for a $5\sigma$ detection with 50\% probability for various burst durations and zenith angle bands for HAWC.}
\label{mu_table}
\end{center}
\end{table}

\begin{figure}[htp]
\centering
\includegraphics[width=0.99\textwidth]{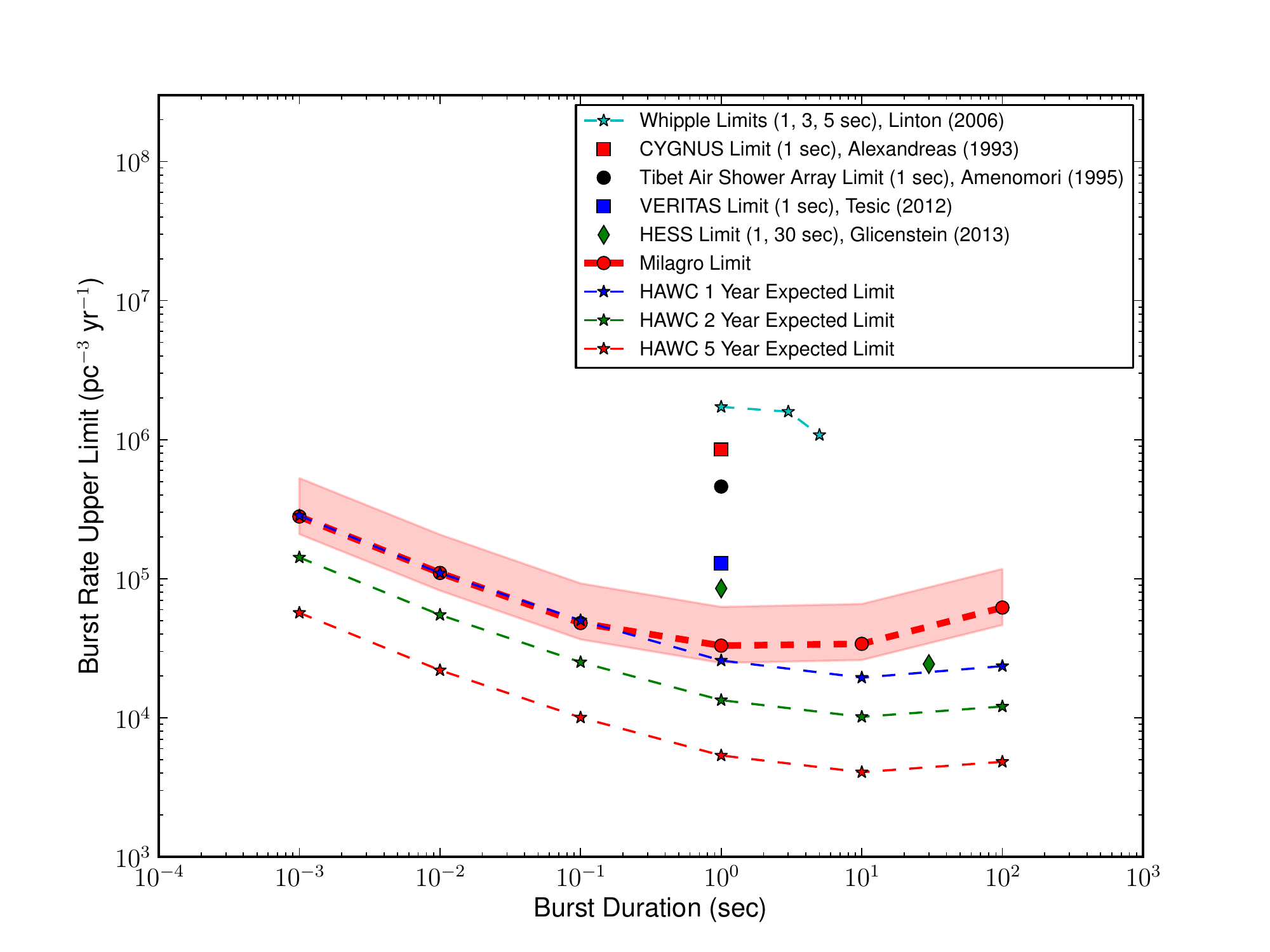}
\caption{PBH Burst Rate Density Upper Limits from Milagro and projected for HAWC, compared with limits from previous direct search experiments~\cite{Amenomori1995,Alexandreas1993,Linton2006,Tesic2012,Glicenstein2013}. The pink
band represents the 50\% systematic uncertainty of the Milagro limit.
All limits are at the 99\% Confidence Level (we have rescaled the reported 95\% CL HESS limit to 99\% CL) and were obtained based on the PBH Standard Emission Model.}\label{pbh_limits}
\end{figure}

The final 99\% confidence level upper limits on the PBH burst rate density are given in Table~\ref{r_table}, assuming zero PBH bursts are observed over the 5 year HAWC search period. For each burst duration, the maximum detectable distance for zenith angle band ${\theta}_1$ and the corresponding effective detectable volume are also shown in Table~\ref{r_table}. We note that HAWC systematic uncertainties have not been included in this study. In Figure~\ref{pbh_limits}, the blue, green, and red thin dashed curves denote the PBH rate density upper limits that HAWC will set if zero PBH bursts
are detected over a one, two and five year search period respectively.
Upper limits based on earlier null detections from various other observatories are also shown in Figure~\ref{pbh_limits}~\cite{Amenomori1995,Alexandreas1993,Linton2006,Tesic2012,Glicenstein2013}. All limits shown in Figure~\ref{pbh_limits} were obtained based on the PBH Standard Emission Model~\cite{MacGibbon1990, Halzen1991}.

\begin{table}[h]
\begin{center}
\scriptsize
\begin{tabular}{|l|c|c|c|}
\hline Burst Duration $\tau$ (s) & $r_{\rm max}$ (pc) & Effective Volume $V(\tau)$ (pc$^3$) & PBH Upper Limit (pc$^{-3}$ yr$^{-1}$) \\ \hline
0.001 & 0.033 & 0.000016 & 56861  \\ \hline
0.01 & 0.044 & 0.000042 & 21976  \\ \hline
0.1 & 0.062 & 0.000092 & 10038  \\ \hline
1.0 & 0.078 & 0.000172 & 5354  \\ \hline
10.0 & 0.089 & 0.000227 & 4059  \\ \hline
100.0 & 0.085 & 0.000191 & 4822  \\ \hline
\end{tabular}
\caption{The maximum detectable distance (for zenith band ${\theta}_1$), the detectable
effective volume and HAWC PBH limit in the event of null detection in 5 years for various remaining PBH lifetimes.}
\label{r_table}
\end{center}
\end{table}

\clearpage

\section{Discussion}

In this work, we report new PBH burst rate density upper limits on parsec scales based on a direct search performed with Milagro data. These new Milagro limits probe various burst timescales previously not investigated. For 1 s bursts, which were probed by numerous earlier experiments, the Milagro limit is now the most constraining. Only the HESS limit at 30 s is more constraining than the Milagro limits. Milagro's successor, the HAWC
observatory, will be even more sensitive to PBH bursts. As seen in Figure~\ref{pbh_limits},
a null detection from a 5 year search with the HAWC Observatory will set PBH upper limits which are significantly better than the upper limits set by any previous PBH burst search including Milagro. Also note that HAWC will surpass the current HESS best limit for a 30 s burst in one year. According to our study, if a PBH explodes within 0.074 parsec (15,000 AU) of Earth and within 26 degrees of zenith, HAWC will have a 95\% probability of detecting it at 5$\sigma$ (as optimized in a 10 s search after trials corrections).
HAWC would see with 95\% probability a PBH burst within 37 degrees of zenith if it happens within 0.058 parsec (12,000 AU) of Earth.

Direct search limits are limits on the local distribution of very low mass black holes, regardless of their initial mass or their formation era and mechanism. A $10^9$ g black hole has a remaining evaporation lifetime of $1$ s and a $7\times 10^{11}$ g black hole has a remaining evaporation lifetime of $5$ years. It can be shown that the evaporation rate~\cite{MacGibbon1991} of an individual black hole whose remaining lifetime is much less than the age of the universe implies that the number density function of any population of BHs with present masses $\sim M<<5 \times 10^{14}$ g has the form $dn/dM\propto M^2$ around $M$~\cite{Carr1976}. From Figure 3, a null detection from a 5 year HAWC search would correspond to an upper limit on the number of local bursts of less than $2.0\times 10^{4}$ pc$^{-3}$ over 5 years and hence to an upper limit on the local density in $7\times 10^{11}$ g or lighter black holes of $\rho(\leq M)\lesssim 5\times 10^{-18}M_\odot$ $\rm{pc}^{-3}$. This applies to very small black holes produced in the present universe as well as primordial black holes. This limit is well below the total (visible and dark matter) local dynamical mass density in the solar neighborhood (the Oort limit) determined from Hipparcos satellite measurements, $\rho_{\odot, Oort} = 0.102(\pm 0.010) M_\odot$ $\rm{pc}^{-3}$~\cite{Holmberg2004}, and the local dark matter density in the solar neighborhood, $\rho_{\odot, DM} = 0.008(\pm 0.003) M_\odot$ $\rm{pc}^{-3}$~\cite{Bovy2012}. If the present number density function for such small black holes, $dn/dM\propto M^2$, can be extrapolated to black holes of present mass $~5\times 10^{14}$ g, a null detection from a 5 year HAWC search would correspond to an upper limit on the local density in $~5\times 10^{14}$ g black holes of $\rho(\leq M)\lesssim 10^{-6}M_\odot$ $\rm{pc}^{-3}$, five orders of magnitude less than $\rho_{\odot, Oort}$.

Direct search limits on the local rate density of PBH bursts are higher than the average cosmological PBH burst rate density limit implied by the 100 MeV extragalactic gamma-ray background constraint on the emission of $~5\times 10^{14}$ g black holes~\cite{PageHawking1976,CarrMacGibbon1998}\footnote{We note that all limits on the expected burst rate derived from the extragalactic or galactic gamma ray or antiproton backgrounds have assumed an extrapolation of the form $dn/dM \propto M^2$ from masses of ~$5\times 10^{14}$ g down to the very small masses of presently bursting PBHs.}. However, PBHs have the gravitational properties of cold dark matter and so should be clustered in our Galaxy, enhancing the local PBH burst rate density by many orders of magnitude over the average cosmological rate~\cite{MacGibbonCarr1991}. Thus a substantial number of PBHs that evaporate producing gamma ray bursts may exist in our Galaxy. PBHs clustered in our Galactic halo should also contribute an anisotropic Galactic gamma-ray background, separable from the extragalactic gamma-ray background. Wright claims that such a halo background has been detected~\cite{Wright1996}. The direct search limits are also weaker than the average Galactic PBH burst rate density limit on kiloparsec scales implied by the Galactic antiproton background~\cite{Abe2012} but the antiproton-derived limit depends on the assumed PBH distribution, the propagation of antiprotons within the Galaxy, and the secondary antiproton component. Direct search limits are independent of assumptions concerning the PBH spatial distribution.

In deriving the new limits, we have assumed the Standard Evaporation Model in which the Hawking-radiated particles leave the vicinity of the black hole without substantially interacting with other Hawking-radiated particles. As shown in detail in \cite{MacGibbon2008}, the energy of the Hawking-radiated particles and the time interval between subsequent emissions are such that self-interactions between the Hawking-radiated particles or their decay products should not form a QED or QCD photosphere around the evaporating black hole nor modify the predicted PBH gamma ray burst spectra. However, if the PBH is embedded in an ambient high density plasma or strong magnetic field, interactions may arise with the net effect that the predicted PBH burst gamma ray spectrum may be enhanced at low energies and decreased at high energies. Because of the remaining BH lifetime's $\tau\propto T^{-3}$ dependence (see Equation~\ref{tempEq}), as-yet unknown particle modes manifesting at high temperatures are unlikely to substantially modify the predicted spectra by more than a factor of $2$. An exception would be the low temperature Hagedorn Model in which all of the final burst emission is produced at a black hole temperature close to the QCD confinement scale, $\Lambda_{QCD} \simeq 250 - 300$ MeV~\cite{PageHawking1976}. The low temperature Hagedorn Model would generate a stronger burst in gamma rays detectable at lower energies but is not consistent with the QCD behaviour observed in terrestrial accelerators~\cite{MacGibbon2008}.

In conclusion, the HAWC observatory has the ability to directly detect emission from nearby PBH bursts.
This capability is scientifically very important, given the large number of
early universe theories that predict PBH formation and the uncertainty in the
degree to which PBHs may cluster locally. A confirmed direct detection of an
evaporating PBH would also provide unparalleled insight into general relativity and high energy particle physics.
\\ \\
\noindent \textbf{Acknowledgments} \\ \\
We gratefully acknowledge Scott DeLay his dedicated efforts in the construction
and maintenance of the HAWC experiment. This work has been supported
by: the National Science Foundation, the US Department of Energy Office
of High-Energy Physics, the LDRD program of Los Alamos National Laboratory,
Consejo Nacional de Ciencia y Tecnologia (grants 55155, 103520, 105033,
105666, 122331 and 132197), Red de Fisica de Altas Energias, DGAPA-UNAM
(grants IN105211, IN108713 and IN121309, IN115409), VIEP-BUAP (grant 161-
EXC-2011), the University of Wisconsin Alumni Research Foundation,
the Luc Binette Foundation UNAM Postdoctoral Fellowship program and the
Institute of Geophysics and Planetary Physics at Los Alamos National Lab.
Many of us are grateful for inspiring discussions with the late Donald Coyne
on the subject of primordial black holes. Finally, we thank the anonymous
referee for comments that significantly improved the paper.

\bibliographystyle{ieeetr}
\bibliography{my_references}

\end{document}